\documentclass[prl,twocolumn,superscriptaddress,showpacs,preprintnumbers,amsmath,amssymb,floatfix]{revtex4}

\usepackage{graphicx}                   
\usepackage{dcolumn}                   
\usepackage{bm}                           
\usepackage{amsmath}
\usepackage{amssymb}
\usepackage{amsfonts}
\begin{document}

\title{Quantum rattling of molecular hydrogen in clathrate hydrate nanocavities
}

\author{L. Ulivi}
\email{lorenzo.ulivi@isc.cnr.it}
\affiliation{ISC--CNR, via Madonna del piano 10, I-50019 Sesto Fiorentino, Italy}
\author{M. Celli}
\affiliation{ISC--CNR, via Madonna del piano 10, I-50019 Sesto Fiorentino, Italy}

\author{A. Giannasi}
\affiliation{ISC--CNR, via Madonna del piano 10, I-50019 Sesto Fiorentino, Italy}

\author{A.J. Ramirez-Cuesta}
\affiliation{Rutherford Appleton Laboratory, ISIS Facility, Chilton, Didcot, Oxon, OX11 0QX, U.K.}

\author{D.J. Bull}
\affiliation{Institute for Materials Research, University of Salford, Salford, Greater Manchester, M5 4WT, U.K.}

\author{M. Zoppi}
\affiliation{ISC--CNR, via Madonna del piano 10, I-50019 Sesto Fiorentino, Italy}
\date{\today}

\begin{abstract}
We have performed high-resolution inelastic neutron scattering studies on three samples of hydrogenated tetrahydrofuran-water clathrates, containing either H$_2$ at different para/ortho concentrtion, or HD.
By a refined analysis of the data, we are able to assign the spectral bands to rotational and center-of-mass translational transitions of either para- or ortho-H$_2$.
The H$_2$ molecule rotates almost freely, while performing a translational motion (rattling) in the nanometric-size cage, resulting a paradigmatic example of quantum dynamics in a non-harmonic potential well.
Both the H$_2$ rotational transition and the fundamental of the rattling transition split into triplets, having different separation.
The splitting is a consequence of a substantial anisotropy of the environment with respect to the orientation of the molecule in the cage, in the first case, or with respect to the center-of-mass position inside the cage, in the second case.
The values of the transition frequencies and band intensities have been quantitatively related to the details of the interaction potential between H$_2$ and the water molecules, with a very good agreement.

\end{abstract}

\pacs{78.70.Nx, 82.75.-z, 63.20.Pw }

\maketitle

In the presence of a low amount of other molecular substances, frozen water can form several crystalline compounds, different from common ice, known as clathrate hydrates \cite{Sloan97}.
The hydrogen-bonded water lattice can host different guest molecules in cages of various sizes and geometries.
The deeply non-harmonic motion of the guest molecules differs from all other normal modes of the crystal, and is responsible for the anomalous thermal conductivity of these materials \cite{Tse05}.
Besides being present in nature, these compounds can be prepared in laboratory and have been recently proposed as effective, safe, and economical materials for hydrogen storage \cite{MaoW02, Florusse04, Lee05}.
Binary hydrogen clathrate hydrates (i.e. made of H$_2$O and H$_2$, only) require $\simeq$ 2000 bar of pressure to be produced at T $\simeq$ 273 K.
However, it has recently been shown that the ternary compound with tetrahydrofuran (THF), can store significant amounts of hydrogen with a much lower formation pressure \cite{Florusse04}.
An understanding of the interaction of the H$_2$ molecule with the host material is fundamental for the rational design of clathrates as hydrogen storage materials.
This can be accessed by an inelastic neutron scattering (INS) study of the dynamics of the molecule trapped in the cage, as presented in this work.

The crystal structure of the H$_2$O--THF--H$_2$ clathrate is cubic (sII structure) with 136 H$_2$O molecules in the unit cell, giving rise to sixteen (small) dodecahedral cages and eight (large) hexakaidecahedral cages \cite{MaoW02, Mak65}.
The THF molecules occupy the larger cages, and no more than one hydrogen molecule is hosted in the small cages \cite{Lokshin04}.
There is an intrinsic advantage in using INS with hydrogenous materials due to the incoherent scattering cross-section of the proton, which is almost two orders of magnitude greater than in any other nucleus, thus allowing for a relatively simple access to the self dynamics of molecular hydrogen.
Hydrogen clathrates for the present experiment were produced at ISC--CNR using D$_2$O and completely deuterated tetrahydrofuran (TDF) in stoichiometric proportion (17:1 mol).
Hydrogen gas at $\simeq 800$ bar was added to the sample by either letting the liquid mixture to freeze in the presence of H$_2$, (at T $\simeq +2 \,^{\circ}\mathrm{C}$) or allowing gaseous H$_2$ to diffuse into the ground D$_2$O--TDF clathrate (at T $\simeq -10 \,^{\circ}\mathrm{C}$).
Gas-release thermodynamic measurements gave results consistent with the hypothesis of single H$_2$ occupancy of the totality of the small cages.
The neutron measurements were carried out on three gas-charged samples and one reference sample of pure D$_2$O--TDF clathrate.
Two samples contained H$_2$ at different ortho-para ({\it o}-H$_2$ - {\it p}-H$_2$)) concentrations, and one sample contained HD.
Due to the procedure of sample preparation, the final {\it o}-H$_2$ concentration of the clathrates obtained starting with {\it normal}-H$_2$ and almost pure {\it p}-H$_2$ resulted 53 \% and 48 \%, respectively ({\it o}-rich sample and {\it p}-rich sample).
Raman measurements performed at ISC--CNR before and after the neutron experiment provided an independent determination of the ortho-para ratio in our hydrogenated samples.

For the present experiment, performed at T=$20$ K on the high resolution ($\Delta \omega/ \omega \lesssim 2.5 \%$) TOSCA spectrometer at ISIS (U.K.), the interesting energy range is between 3.5 and 120 meV.
Many features due to acoustic, optic, and molecular excitations of the lattice were recognizable in the spectrum of the reference sample.
The spectra from the hydrogen-charged clathrates showed, in addition, the presence of several narrow and intense bands due to the molecular hydrogen dynamics.
The hydrogen contribution was obtained by subtraction of the reference spectrum, which has a low intensity ( $5 \div 15$ \% of the main bands peaks).
It was possible to interpret the observed lines unambiguously as being due to the molecular rotation (E $\simeq$ 14.7 meV), to the fundamental transition of the quantum rattling (E $\simeq$ 10.0 meV), and to their combinations, as discussed below.

Molecular hydrogen trapped in the clathrate cages appears, in general, as a non-equilibrium mixture of {\it o}-H$_2$ and {\it p}-H$_2$.
Thus, we need to consider the neutron scattering cross section for the two species, and its dependence on the rotational transitions \cite{Lovesey84}.
Neglecting the coherent part of the scattering, on account of the overwhelming incoherent scattering length for the proton, the double differential neutron scattering cross section is proportional to the self-part of the dynamical structure factor for the motion of the center of mass (c.o.m.) $S_{\mathrm{self}} (Q,\omega)$ and is written as \cite{Colognesi04}
\begin{equation}
\frac{d^2 \sigma}{d\Omega d\omega}=\frac{ k_\mathrm{f}}{ k_\mathrm{i}}S_{\mathrm{self}} (Q,\omega) \otimes \sum_{JJ'}\delta(\omega-\omega_{JJ'})\nu(J,J',Q),
\end{equation}
where $k_\mathrm{i}$ and $k_\mathrm{f}$ denote the initial and final neutron wavevectors, $\omega=E/\hbar$, and the symbol $\otimes$ denotes a convolution product.
This expression holds if the hypothesis of decoupling of the internal (rotational) motion of the molecule from the c.o.m. motion is satisfied.
The Dirac $\delta$-functions are centered at the energies of the rotational transition, $\hbar \omega_{JJ'}$ of the H$_2$ molecule and the intensity factor $\nu(J,J',Q)$ is a function of the momentum transfer $Q$ and depends on the rotational transition $J \!\! \rightarrow \!\! J'$ of the molecule.
The expected spectrum therefore consists of the c.o.m. excitations replicated and shifted by the energy of any possible rotational excitation of the molecule.
Since INS is not subject to selection rules, all transitions to molecular rotation and c.o.m. vibration states are allowed.
The intensity factors $\nu(J,J',Q)$ have been calculated assuming a rigid rotor molecule \cite{Young64}.
At the low temperature of the experiment, only the lowest rotational states (i.e. $J\!=\!0,1$) are populated, and few transitions contribute to the spectrum in the observed frequency region, namely the rotationally-elastic $J \!\! = \!\! 1 \!\! \rightarrow \!\! 1$ and the inelastic $J \!\! = \!\! 1 \!\!  \rightarrow \!\! 2$ transition of {\it o}-H$_2$ and the inelastic $ J \!\!=\!\! 0 \!\!\rightarrow \!\! 1 $ of {\it p}-H$_2$. 
We would note that the $J \!\! = \!\! 0 \!\! \rightarrow  \!\! 0 $ transition of the {\it p}-H$_2$ molecule, being weighted by the coherent scattering length only, does not contribute appreciably to the spectrum and, consequently, the observed fundamental rattling transition (E $\simeq$ 10.0 meV) is mainly due to the ortho molecules.

%
\begin{figure}[thb!]
\includegraphics[bb= 0.5cm 1.0cm 19cm 27cm, width=6.0cm]{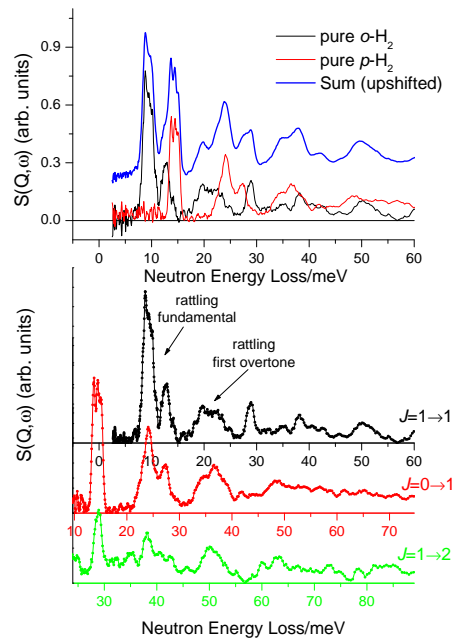}
\caption{(Color online) Upper panel: The clathrate spectrum (blue line, vertically displaced) can be experimentally decomposed into the sum of the {\it o}-H$_2$ and {\it p}-H$_2$ contributions (black and red line respectively).
The spectrum of {\it p}-H$_2$ shows the strong $ J \!\! = \!\! 0 \!\! \rightarrow \!\! 1 $ rotational band (a triplet at $\simeq$ 14 meV), while the {\it o}-H$_2$ spectrum displays the (split) band around 10 meV, due to the quantum rattling motion of the molecule in the cage.
In the lower panel the {\it p}-H$_2$ (red) and {\it o}-H$_2$ spectra (green) have been redrawn shifted by $-\Delta E_{01}= -14.7$ meV and by $-\Delta E_{12}= -29.0 $ meV, respectively, to show the coincidence, in frequency and shape, of the rotational--rattling combination bands with the rattling fundamental and first overtone.
\label{f.sptstrasl} }
\end{figure}
For unambiguous assignment of all of the spectral features, we compare the spectra of the {\it o}-rich and {\it p}-rich samples.
The intensity ratio of the two bands at $\simeq$10 meV and $\simeq$14 meV is {\em lower} in the {\it p}-rich sample, confirming that the 10 meV band is due to {\it o}-H$_2$ (namely, to the combination of the {\em elastic} $J \!\! = \!\! 1 \!\! \rightarrow \!\! 1 $ rotational transition with the fundamental rattling transition of the c.o.m. motion), and the 14 meV band to {\it p}-H$_2$ (namely, to the {\em inelastic} $J \!\! = \!\! 0 \!\! \rightarrow \!\! 1$ rotational transition).
By combining linearly the {\it p}-rich and {\it o}-rich spectra, we are able to obtain the spectra of pure {\it p}-H$_2$ and pure {\it o}-H$_2$.
The result is particularly elucidating and is shown in Fig.~1.
The spectrum of {\it o}-H$_2$ (black line) contains contributions from the {\em fundamental} transition of {\em rattling } vibration, and from its {\em overtones}, while the {\it p}-H$_2$ spectrum (red line) contains the pure rotational ($J \!\! = \!\! 0 \!\! \rightarrow \!\! 1$) transition at about 14 meV plus {\em combinations} of this one with the c.o.m. transitions.
This can be demonstrated just by shifting the {\it p}-H$_2$ spectrum by the energy of the rotational transition (red line in the lower panel of Fig.~1) obtaining an almost perfect match of the translational bands.
The {\it o}-H$_2$ spectrum contains also the band due to the $J \!\! = \!\! 1 \!\! \rightarrow \!\! 2$ transition (about 29 meV) and the combination of this one with the c.o.m. spectrum (shifted spectrum, green line in the lower panel of Fig.~1).

%
\begin{figure}[t!]
\includegraphics[bb= 0.5cm 6.5cm 19.5cm 21cm, width=8.5cm]{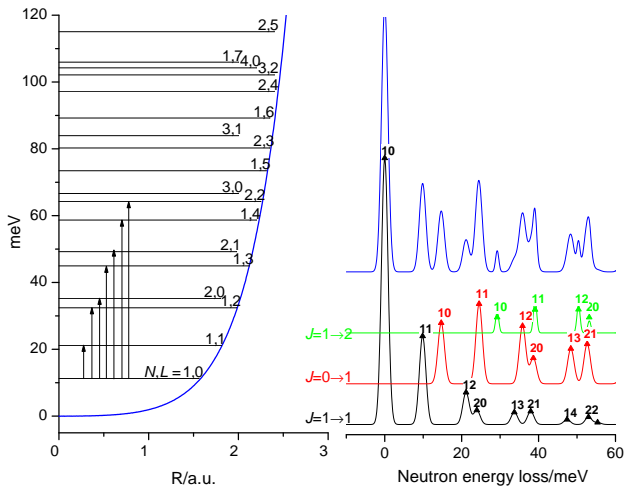}
\caption{(Color online) Left panel: Isotropic part of the potential energy for one H$_2$ molecule as a function of the distance $R$ (atomic units) from the center of the dodecahedral cage, and calculated energy levels.
Right panel: Calculated incoherent INS spectra for one H$_2$ molecule (Eq.~1).
The spectrum (blue line) is the sum of three contributions, where the transition lines of the c.o.m. motion (labeled by the final values of $N,L$) are shifted by the amount $\Delta_{JJ'}$ of each rotational transition: $\Delta_{11}= 0$ (black line), $\Delta_{01}= 14.7 $ meV (red line), $\Delta_{12}= 29.0 $ meV (green line).
\label{f.potlevels} }
\end{figure}
\begin{figure}[hbt!]
\includegraphics[bb= 0.0cm 2.0cm 18cm 28cm, width=6.0cm]{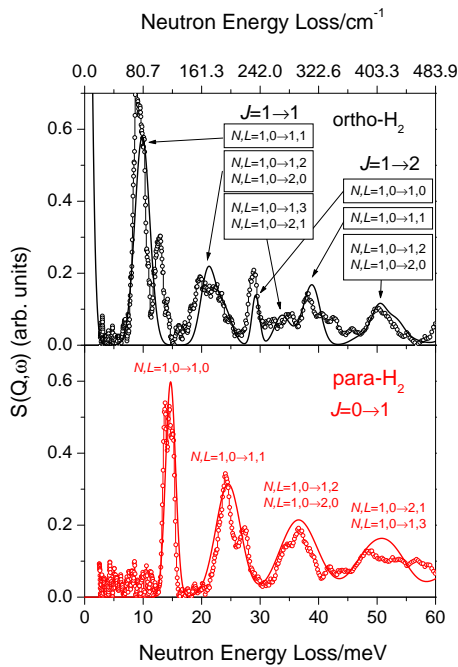}
%
\caption{(Color online) Comparison of the {\it o}-H$_2$ and {\it p}-H$_2$ experimental spectra with the results of our calculation.
We have considered the $J \!\! = \!\! 1 \!\! \rightarrow \!\! 1$ and $J \!\! = \!\! 1 \!\! \rightarrow \!\! 2$ molecular rotational transitions for the o-H$_2$ spectrum and the $J \!\! = \!\! 0 \!\! \rightarrow \!\! 1$ transition for the p-H$_2$ spectrum.
The various bands are assigned.
\label{f.sptsassegnati} }
\end{figure}
%
%
For comparison with the experimental determination, we have calculated the spectral intensity of the rattling excitations by solving numerically the Schr\"odinger equation for the c.o.m. motion.
The potential energy for one H$_2$ molecule, as a function of the c.o.m. displacement $ \vec R $ from the center of the dodecahedral cage, has been calculated summing the pair H$_2$--D$_2$O potential (assumed equal to the H$_2$--H$_2$O potential \cite{Alavi06,Hodges04}) over 514 molecules of D$_2$O around the center of the cage on a rigid lattice, and averaging over H$_2$ molecular orientations.
The disorder of the water deuterons has been taken into account performing several averages over random configurations, all of them respecting the Bernal and Fowler ice rules \cite{Bernal33}.
A further average over the direction of $\vec{R} $ gives the isotropic part of the potential.
The potential energy of the hydrogen molecule as a function of $ R= | \vec{R}| $, obtained using the two potential models available from literature \cite{Alavi06,Hodges04}, coincide for the purpose of this analysis, and is represented in Fig.~2.
It is strongly anharmonic and rather flat in the center.
As a consequence, the calculated energy levels, (labeled by the principal quantum number $N=1, 2, 3, ...$), are split (the energy depends on the orbital quantum number $L=0, 1, 2, ...$) and not equally spaced.
The residual ($2L+1$)--fold degeneracy of each level is not removed in this model due to the assumed isotropy of the potential.
The knowledge of the wave function for the translation degrees of freedom, and the use of equation (1), where all coefficients $ \nu(J,J',Q) $ are known, permits the calculation of the intensity of transitions, determining the whole neutron energy loss spectrum.
For each rotational transition $J \!\!  rightarrow \!\! J'$ of the molecule, we obtain a spectrum given by the rattling transition energies plus the rotational energy transition  $\Delta E_{JJ'}$.
The calculated spectra are shown in Fig.~2, where all transitions are represented by Gaussian lines having the same (arbitrary) width.
In Fig.~3, we compare separately the {\it o}-H$_2$ and {\it p}-H$_2$ experimental spectra with the results of our isotropic model, finding a quantitatively good level of agreement, taking into account that this simple model cannot reproduce the splitting of the bands.

The assignment of the bands is confirmed by the results obtained for the HD--clathrate, compared to the spectrum of the H$_2$--clathrate in Fig.~4.
We also observe that the experimental band due to the {\em fundamental} of the rattling mode ($\simeq$ 10 meV in the H$_2$ spectrum) is split in more than two components, evidencing an appreciable anisotropy of the c.o.m. potential in the cage.
In contrast, the splitting of the $ J \!\! = \!\! 0 \!\! \rightarrow \!\! 1$ {\em rotational} transition into a triplet ($\simeq$ 14 meV in the H$_2$ spectrum) is a consequence of the anisotropy with respect to the {\em orientation} of the H$_2$ molecule.
The cage shape, as it results from the structural measurements \cite{Mak65,Jones03}, is quite anisotropic, with the oxygen atoms located at three different distances from the center.
Fitting each band with three Voigt lineshapes (with a Gaussian width fixed by the instrumental resolution), we obtain the experimental values reported in table 1.
Comparing the (average) experimental frequencies with the results of our model we obtain a very good agreement.
In particular, the ratio of the rattling frequencies measured for H$_2$ and HD is about 1.40, and is well reproduced by our calculation.
Note that this value is intermediate between the harmonic value $ \sqrt{m_{\mathrm{HD}} / m_{\mathrm{H_2}}}=\sqrt{3/2}=1.225 $ and the value expected for a square well potential  $ m_{\mathrm{HD}}/ m_{\mathrm{H_2}} =3/2=1.5 $. 

Recently the energy levels of one H$_2$ molecule confined inside an isolated dodecahedral cage of 20 H$_2$O molecules, (similar for dimensions and shape to the one of the THF--H$_2$O clathrate), has been calculated, solving numerically the five-dimensional Schr\"odinger equation, and taking into account the anisotropy of the potential energy and the coupling of rotational and translational motion \cite{Xu06}.
The authors predicted a splitting into a triplet of both the rattling fundamental and the rotational transition, as we have experimentally observed (see Table~1).
\begin{table}[t!]
\begin{center}
\begin{tabular}{|c|ccc|cc|}
\hline\hline
    & \multicolumn{3}{c|} {H$_2$} & \multicolumn{2}{c|} {HD} \\ 
    & \text{exp.} & \text{model} & \text{ref.~\protect{\onlinecite{Xu06}} } & \text{exp} & \text{model} \\ 
\hline 
     \multicolumn{6}{|c|} {Rattling fundamental}  \\ 
\text{I}       &    8.80        &   &      6.39 & 6.59 &                  \\
\text{II}       &    9.94         &   &      8.38 & 7.28 &                   \\
\text{III}    &    12.53         &   &      9.91 & 8.75 &                   \\
\text{ave.}      &    9.86       & 9.85   &      8.22 & 7.04 &     7.44              \\
\hline 
     \multicolumn{6}{|c|} {Rotation}  \\ 
\text{I}              &   13.64         &   &      11.32 & 10.86 &                   \\
\text{II}           &    14.44         &   &      14.66 & 11.58 &                   \\
\text{III}           &    15.14         &   &      18.83 & 12.21 &                   \\
\text{ave.}          &    14.41         &   &      14.93 & 11.54 &                   \\
\hline\hline
\end{tabular}
\end{center}
\caption{\label{t.energies} Experimental and calculated transition energies (meV).
Roman numbers indicate the different components of each band, and ``ave.'' means weighted average.
}
\end{table}
%
\begin{figure}[b!]
\includegraphics[bb= 2cm 2.0cm 19cm 27cm, width=6.5cm]{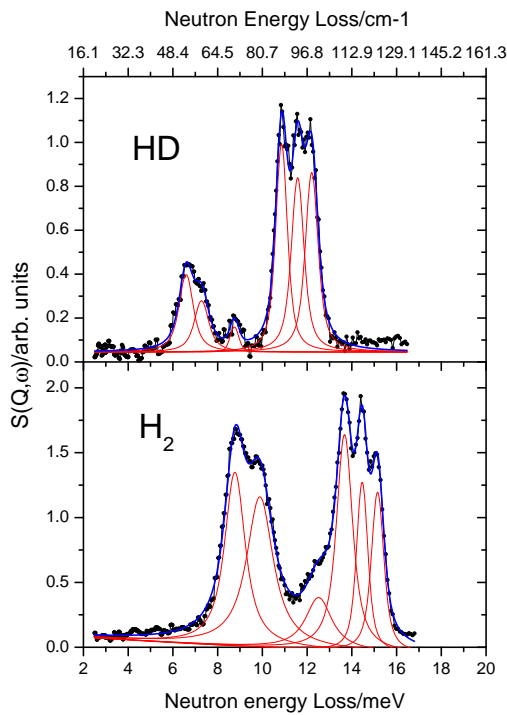}
%
\caption{(Color online) Rotational and fundamental rattling transition for a H$_2$ (lower panel) and HD (upper panel) molecule in a THF-clathrate.
Each band is fitted with a triplet (Voigt shape) whose frequencies are reported in Table 1.
\label{f.H2andHD} }
\end{figure}
While the calculated splitting for the rattling transition (3.52 meV) reproduces correctly the experimental one (3.73 meV), the calculation strongly overestimates the splitting of the rotational transition (7.51 {\it vs} 1.50 meV).
In addition, for the fundamental transition of the rattling mode, our calculated result is somewhat more similar to the experimental value, probably because we used a much larger number of molecules, though using the simplification of an isotropic potential.
Then, therefore, the pair potential model used in Ref.~\onlinecite{Xu06} seems to overestimate the actual anisotropic forces on the hydrogen molecule.

In summary, we have measured the quantum dynamics of a single H$_2$ molecule in the confined geometry of a water clathrate nanocavity.
An isotropic model describes quantitatively the main features of the motion, were coupling between rotational and c.o.m. dynamics are weak, reproducing frequencies and intensities of the observed spectrum.
The splitting of the rotational and translational bands is a consequence of the anisotropy of the environment, which must be modeled with greater accuracy than done until now, to extract direct information on the basic interaction between H$_2$ and H$_2$O molecules.
\begin{acknowledgments}
The Cooperation Agreement No.01/9001 between CNR and CCLRC, the Grant from Firenze-Hydrolab by the Ente Cassa di Risparmio di Firenze and the E.U. contract MRTN-CT-2004-512443 HYTRAIN are gratefully acknowledged.
\end{acknowledgments}

\end{document}